%% file: main.tex
\documentclass[ALICE,manyauthors]{cernphprep}
\usepackage[comma,square,numbers,sort&compress]{natbib}
\usepackage{hyperref}
\usepackage{lineno}
\usepackage{xspace}
\usepackage{comment}
\usepackage{xcolor}
\usepackage[T1]{fontenc}
\usepackage{orcidlink}

\begin{document}
\input{commands.tex}

\begin{titlepage}
\PHyear{2025}       
\PHnumber{070}      
\PHdate{26 March}  

\title{Multiplicity-dependent inclusive \jpsi production at forward rapidity in pp collisions at $\mathbf{\s= 13}$~TeV}

\ShortTitle{Multiplicity-dependent \jpsi production at forward rapidity in pp collisions}

\Collaboration{ALICE Collaboration\thanks{See Appendix~\ref{app:collab} for the list of collaboration members}}
\ShortAuthor{ALICE Collaboration} 

\begin{abstract}
This paper presents a study of the inclusive forward $J/\psi$ yield as a function of forward charged-particle multiplicity in pp collisions at $\s=13$~TeV using data collected by the ALICE experiment at the CERN LHC. The results are presented in terms of relative $J/\psi$ yields and relative charged-particle multiplicities with respect to these quantities obtained in inelastic collisions having at least one charged particle in the pseudorapidity range $|\eta| < 1$. The $J/\psi$ mesons are reconstructed via their decay into $\mu^+ \mu^-$ pairs in the forward rapidity region ($2.5 < y < 4$). The relative multiplicity is estimated in the forward pseudorapidity range which overlaps with the \jpsi rapidity region. The results show a steeper-than-linear increase of the $J/\psi$ yields versus the multiplicity. They are compared with previous measurements and theoretical model calculations.

\end{abstract}
\end{titlepage}

\setcounter{page}{2} 


\section{Introduction}
\label{sec:intro}
The production of charmonia, bound states of charm and anti-charm quarks ($c\bar{c}$), in high-energy hadronic collisions is not yet fully understood and can be considered as a valuable probe for the theory of strong interactions -- the Quantum Chromodynamics (QCD). Charmonium production can be factorized into two distinct processes, namely the hard-scale and soft-scale processes. At the initial state of high-energy hadronic collisions, $c\bar{c}$ pairs are produced via processes involving large momentum transfer (hard processes), which can be well described with perturbative QCD (pQCD) approaches. The subsequent evolution and hadronization of these pairs into a colorless bound charmonium state is a soft-scale process, which can be addressed via phenomenological model calculations. Calculations based on non-relativistic QCD (NRQCD) provide a satisfactory description of the \jpsi production spectrum and its rapidity dependence~\cite{Ma:2014mri} as well as the \jpsi polarization~\cite{Ma:2018qvc} in high-energy proton-proton (pp) collisions. Despite the overall success of these calculations, significant deviations with respect to the measured \jpsi production spectrum at high transverse momenta (above few tens of GeV/$c$) are to be noted~\cite{ATLAS:2023qnh}.
The NRQCD-based calculations rely on the factorization approach discussed above. The initial state of the incoming protons is described within the Color Glass Condensate (CGC) framework~\cite{Gelis:2010nm}. The transition from the intermediate $c\bar{c}$ state to colorless charmonium state is implemented via universal and non-perturbative long-distance matrix elements (LDMEs) extracted from experimental data. Both color-singlet and color-octet intermediate states are considered.
Within the CGC approach, the multiplicity dependence of the \jpsi production is in general steeper-than-linear~\cite{Ma:2018bax}. The reason is in the Bjorken-$x$ dependence of the saturation scale that controls the parton saturation in the incoming protons. The scale increases with decreasing $x$, therefore the particle multiplicity, produced mainly by low-$x$ partons, is suppressed more compared to the \jpsi production, which originates from partons having higher values of $x$. In Refs.~\cite{Levin:2019fvb,Gotsman:2020ubn}, it is argued that 3-Pomeron fusion significantly enhances the production of color-singlet intermediate state and thus drives charmonium production. Measurements of the \jpsi production as a function of the particle multiplicity can shed more light on the role of the hard and soft processes involved in the charmonium production and further constrain the theoretical models.

Various theoretical models and event generators, incorporating different mechanisms to produce \jpsi and charged particles, have been used to understand the experimental results. Within the PYTHIA generator framework~\cite{Sjostrand:2014zea}, the produced particle multiplicity in pp collisions at TeV energies is driven to a large extent by the multi-parton interactions (MPI) mechanism. The MPIs are semi-hard scatterings of partons from the incoming protons and their number ($N_{\rm MPI}$) significantly varies collision by collision~\cite{Weber:2018ddv}. Up to about 3 times its mean value, the $N_{\rm MPI}$ is determined by the amount of the geometrical overlap between the colliding protons. Beyond, in the most central pp collisions, the $N_{\rm MPI}$ is driven by statistical fluctuations. Within the PYTHIA framework, \jpsi are produced from several sources~\cite{Weber:2018ddv}. The first source is the perturbative scattering process implemented within the NRQCD model, via color-singlet and color-octet intermediate bound states. The second source, called cluster collapse, is a binding at the hadronization stage of charm and anti-charm quarks that are close in phase space. The third source is the production of non-prompt \jpsi via weak decays of beauty hadrons. In a baseline scenario, the number of produced \jpsi and the particle multiplicity are both essentially proportional to $N_{\rm MPI}$. However, the hadronization, implemented in PYTHIA via the Lund string model, is affected by the so-called Color Reconnection (CR) mechanism. The hadronizing strings can be rearranged through CR to reduce the total string tension. This results in a lower final-state particle multiplicity. In addition, during the hadronization stage, the CR mechanism allows \jpsi production via the binding of charm and anti-charm quarks coming from two different hard scattering processes. This results in a quadratic dependence of the number of produced \jpsi via cluster collapse as a function of $N_{\rm MPI}$.

The dependence of the \jpsi production on the particle multiplicity can also be affected by the so-called autocorrelation effects. The charged-particle multiplicity and \jpsi yield are auto-correlated when the \jpsi production leads to an increase of the charged-particle multiplicity, as described by various processes in Ref.~\cite{Weber:2018ddv}. For example, non-prompt \jpsi is produced together with particles from the beauty-quark fragmentation and the decay of the corresponding beauty hadron. As a consequence, in case the \jpsi production and the particle multiplicity are measured in the same rapidity range, the non-prompt \jpsi yields are biased towards steeper-than-linear dependence on the multiplicity. 
Autocorrelation effects can also be present if \jpsi are produced via the NRQCD and cluster collapse mechanisms due to the production of associated particles. Finally, another autocorrelation effect arises due to the presence of \jpsi decay daughters. Experimentally, the \jpsi are reconstructed via their dilepton decay mode. In case the contribution of the two leptons is not subtracted from the measured multiplicity, the observed \jpsi yields would show a steeper dependence on the multiplicity.

In practice, the measurements of \jpsi production as a function of the multiplicity can be done in various ways. The \jpsi production yield and the charged-particle multiplicity can be measured at midrapidity or forward rapidity, with the rapidity ranges overlapping or being distinct. The multiplicity estimator, which defines the event classes in which the charged-particle multiplicities and \jpsi yields are measured, can be obtained by counting charged particles within a rapidity range that coincides with that of the \jpsi, that of the charged-particle multiplicity, or both. The role of the multiplicity estimator is essential, since it can significantly affect the values of both the charged-particle multiplicities and the \jpsi production yields observed in the various event classes. In high-energy pp collisions, the correlation between the multiplicities in two distinct pseudorapidity ranges increases with a decreasing pseudorapidity gap between the ranges~\cite{ALICE:2020swj}. The correlation ultimately reaches 100\% when the two ranges fully overlap. Therefore, given the steeply falling shape of the multiplicity, the spanned range of the measured multiplicity increases when the gap between the pseudorapidity ranges of the multiplicity estimator and the multiplicity measurement decreases~\cite{ALICE:2020swj}. The spanned range becomes widest when the two ranges fully overlap. The multiplicity estimator can directly affect the observed \jpsi yields as well. For example, in case \jpsi are produced together with associated particles, using overlapping (pseudo)rapidity ranges for the multiplicity estimator and the measurement of \jpsi yields leads a systematic migration of the collisions containing the \jpsi towards higher multiplicity event classes and thus to a steeper dependence of the \jpsi yields on the charged-particle multiplicity. Autocorrelation effects like the ones described above can affect the measurements to various degrees. This has to be taken into account when interpreting the results and, in general, requires detailed comparison with event generators and theoretical models in which the \jpsi yields as a function of the charged-particle multiplicity are obtained in the same way as in the experimental data.

The ALICE Collaboration~\cite{ALICE:2022wpn} at the LHC has measured both mid and forward rapidity \jpsi yields as a function of the charged-particle multiplicity in pp collisions~\cite{ALICE:2012pet,ResultsMidrap,ResultsForward}. In the midrapidity case, the charged-particle multiplicity is also measured at midrapidity within a range that is largely overlapping with that of the \jpsi~\cite{ResultsMidrap}. The multiplicity selection is performed using two different estimators based on mid and forward rapidity charged-particle multiplicities. In both cases, the midrapidity \jpsi yields show a steeper-than-linear dependence on the multiplicity, with the only observed difference being the narrower multiplicity and \jpsi yield ranges spanned with the forward rapidity estimator. The measurement at forward rapidity is performed at various collision energies~\cite{ALICE:2012pet,ResultsForward}. Midrapidity charged particles are used both for the multiplicity selection and the multiplicity measurement. In contrast to the midrapidity case, the observed dependence of the \jpsi yields as a function of multiplicity is close-to-linear, with little to no dependence on the collision energy. It is also worth mentioning the measurements by the STAR~\cite{STAR:2018smh} and PHENIX~\cite{PHENIX:2024dqs} Collaborations at RHIC in pp collisions at a significantly lower energy of $\s=200$~GeV. Strong autocorrelation effects related to the presence of \jpsi decay daughters are observed. In case the contribution of the daughters is subtracted or the multiplicity is estimated in a rapidity range that does not overlap with that of the \jpsi, the \jpsi yields versus the charged-particle multiplicity show a significantly less-than-linear increase~\cite{PHENIX:2024dqs}. A similar autocorrelation effect can be present in the ALICE midrapidity measurement, although the significantly larger event multiplicity at the LHC energy should make the relative contribution of the \jpsi decay products much less sizable. This is somehow indicated by the similarity of the results obtained with mid and forward rapidity multiplicity estimators in Ref.~\cite{ResultsMidrap}, although the direct comparison between these two results is difficult. The measurement using the forward multiplicity estimator is presumably less affected by autocorrelation effects and therefore the expected multiplicity dependence of the \jpsi yields is less steep~\cite{Weber:2018ddv}. At the same time, the rapidity gap between the multiplicity measurement and the multiplicity estimator reduces their correlation. This leads to a reduced spanned range of the measured multiplicity and hence to a steeper multiplicity dependence of the \jpsi yields.

The purpose of the presented analysis is the measurement of the forward (2.5 $<$ $y$ $<$ 4) \jpsi yields as a function of the forward (-3.7 $<$ $\eta$ $<$ -1.7) charged-particle multiplicity in pp collisions at $\sqrt{s}=13$ TeV. The \jpsi are reconstructed via their dimuon decay mode. In contrast to the previous measurements at forward rapidity, the chosen pseudorapidity range for the multiplicity selection and measurement largely overlaps with the \jpsi rapidity range. It is essential to note that the ALICE experiment uses different conventions for pseudorapidity $\eta$ and rapidity $y$: $\eta$ is given in the laboratory frame and the muons are reconstructed in -4 $<$ $\eta$ $<$ -2.5, whereas the corresponding \jpsi are referred to as forward and having 2.5 $<$ $y$ $<$ 4. In other words, even though the reconstructed \jpsi have negative rapidity in the laboratory frame, we flip the sign of their rapidity and refer to them as having positive rapidity in order to comply with the convention used in numerous quarkonium measurements performed by ALICE.
The contribution of the two decay muons is subtracted from the multiplicity estimator. The measurement is done for inclusive \jpsi, i.e., including the contribution from non-prompt \jpsi, because the data used in the present analysis are taken with an ALICE detector layout not designed to separate prompt and non-prompt \jpsi in the forward direction. The results are presented in terms of relative \jpsi yields and relative charged-particle multiplicities with respect to these quantities obtained in 'INEL$>$0' collisions - inelastic collisions having at least one charged particle in the range $\lvert \eta \rvert <1$~\cite{ALICE:2015olq}. The main goal is to study the rapidity dependence of the \jpsi production mechanisms and the role of possible autocorrelation effects by comparing the results to the previous measurements at mid and forward rapidity and state-of-the-art particle production models.

\section{Detectors and data samples}
A detailed description of the ALICE apparatus can be found in Refs.~\cite{ALICE:2008ngc,ALICE:2014sbx}. In the following, only the detectors which are essential for the present analysis are described.

The Silicon Pixel Detector (SPD) is the innermost detector of the ALICE central barrel. It consists of two cylindrical layers, covering pseudorapidity ranges $|\eta|$ $<$ 2 and $|\eta|$ $<$ 1.4, respectively. In the present analysis, the detector is mainly used for the reconstruction of the primary collision vertex.

The V0 detector is composed of two arrays of scintillator counters positioned on both sides of the interaction point - the V0A detector covers the pseudorapidity range 2.8 $<$ $\eta$ $<$ 5.1 and the V0C detector covers the pseudorapidity range -3.7 $<$ $\eta$ $<$ -1.7~\cite{ALICE:2013axi}. Each array consists of 4 rings and each ring is divided into 8 sectors. The V0 detector measures the energy deposited by charged particles in its scintillator counters. The V0 detector provides the minimum-bias (MB) trigger which requires a coincidence of signals in both V0A and V0C. Together with the SPD, the V0 is used for the removal of the beam-induced background and pileup collisions. In the present analysis, the V0C detector is used as a multiplicity estimator.

The Muon Spectrometer (MS) is used to reconstruct the muons from the \jpsi decays within the pseudorapidity range -4 $<$ $\eta$ $<$ -2.5. The MS is composed of a front absorber of 10 interaction lengths followed by 5 tracking stations. The third station is located inside a dipole magnet with a 3~Tm magnetic-field integral. The tracking stations are followed by a 7.2 interaction length iron wall and two trigger stations. Single muon triggers are formed by a coincidence of the MB trigger and the presence of at least one track segment reconstructed in the muon trigger stations, while dimuon triggers are formed by a coincidence of the MB trigger and a pair of track segments in the muon trigger stations. Each track segment is required to have a transverse momentum above 1 GeV/$c$. The corresponding muon trigger efficiency reaches 50\% at this threshold value and saturates, close to unity, for transverse momenta above 2 GeV/$c$.

The analysis is based on pp collision data collected by the ALICE experiment during the LHC Run 2 between 2016 and 2018. Two main data samples are employed: 1.17 billion minimum-bias events for the multiplicity measurement and the definition of the multiplicity classes, and dimuon-triggered data corresponding to about 25~pb$^{-1}$ integrated luminosity for the measurement of the \jpsi production yields.

\section{Analysis}
The analysis is mainly based on approaches used in previous measurements performed with the ALICE detector~\cite{ResultsMidrap, ResultsForward, ALICE:2022gpu}. In the following, only the points specific to the present analysis will be discussed in more details.

\subsection{Event, muon and dimuon selection}
\label{sec:selection}

The event, muon and dimuon selection criteria are equivalent to the one employed in Refs.~\cite{ResultsForward,ALICE:2022gpu}. The selected events are required to have a primary vertex, reconstructed by the SPD, passing vertex quality selection and having a longitudinal coordinate $z_{\rm vtx}$ within 10~cm from the nominal interaction point. The beam-induced background is removed by requiring that the timing of the signals in both V0A and V0C detectors is compatible with beam-beam collisions. The pileup collisions are rejected based on the correlations between the total V0 signal, the number of SPD hits and the number of SPD tracklets.

The muon tracks are required to be within the pseudorapidity range -4 $<$ $\eta$ $<$ -2.5. The radial distance of the tracks to the $z$ axis at the end of the absorber is required to be between 17.6 and 89.5 cm. Only tracks that match corresponding track segments in the trigger stations are accepted. Background tracks are removed with a selection on the product of the total track momentum and the distance of closest approach to the primary vertex in the transverse plane. Only pairs of opposite-sign muon tracks are considered. The dimuons are required to be in the rapidity range 2.5 $<$ $y$ $<$ 4 and to have an invariant mass $m_{\mu\mu}$ between 2 and 5 GeV/$c^2$. 

\subsection{Charged-particle multiplicity selection}
\label{sec:mult_sel}

\begin{figure}[htb!]
    \centering
    \includegraphics[width=0.8\linewidth]{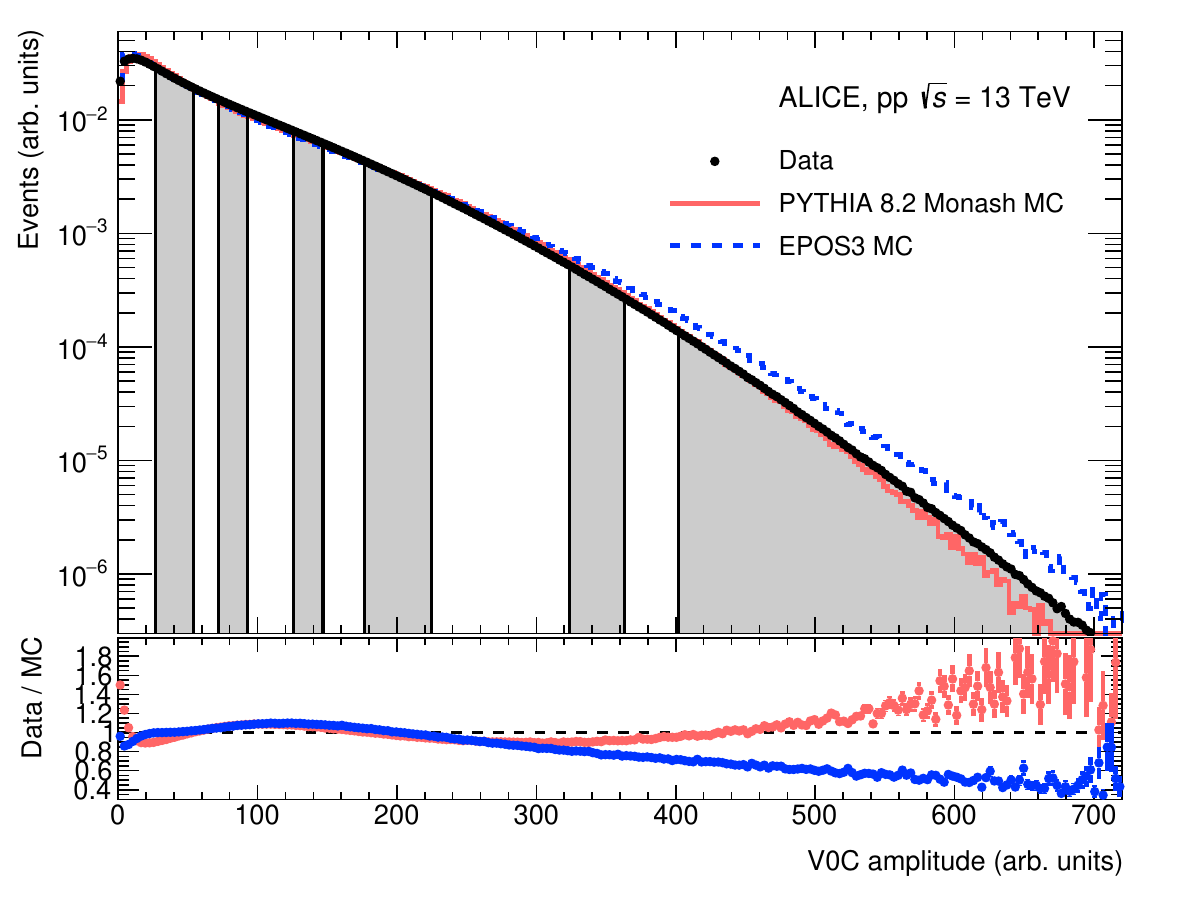}
    \caption{The V0C distributions in minimum-bias events in data, PYTHIA~8.2 Monash MC~\cite{Sjostrand:2014zea} and EPOS3 MC~\cite{Werner:2013tya}. The chosen multiplicity classes are indicated by alternating gray and white areas. The percentile ranges corresponding to the multiplicity classes (from low to high multiplicities) are 70-100\%, 50-70\%, 40-50\%, 30-40\%, 20-30\%, 15-20\%, 10-15\%, 5-10\%, 1-5\%, 0.5-1\%, 0.25-0.5\%, 0-0.25\%. The bottom panel shows the ratio between data and MC.}
    \label{fig:V0CPythiaData}
\end{figure}
The charged-particle multiplicity selection is entirely based on the signal recorded by the V0C detector. As a first step, the dependence of the V0C acceptance on the longitudinal vertex position is corrected event-by-event based on the mean V0C signal in bins of $z_{\rm vtx}$. This is done separately for each data-taking run. As a second step, the V0C signal is corrected for detector aging. In principle, the detector aging is taken care of by a calibration procedure which is performed at the beginning of each data-taking year. Nevertheless, residual aging effects are present within each year and have to be taken into account in the analysis. The aging correction is obtained by comparing the mean V0C signal in each run to a global reference value and is also applied event-by-event. The values of the correction range within $\pm$ 25\%. The corrected V0C distributions obtained in various data-taking periods are found to be fully compatible without any deviations beyond the expected statistical fluctuations, meaning that there is no sizable non-uniformity of the detector aging and that the data from all data-taking periods can be merged.  

The multiplicity classes are obtained using the corrected V0C signal and are defined in terms of the percentile of the minimum-bias data sample. The V0C signal distribution and the chosen multiplicity classes are shown in Fig.~\ref{fig:V0CPythiaData}. The data, both MB and dimuon-triggered samples, are then split according to the defined multiplicity classes. The measurements of the charged-particle multiplicity and the \jpsi yields are performed individually in each class.

In the case of events containing dimuons, the contribution of the two muons is subtracted beforehand from the measured V0C signal. Data collected with the single muon trigger are used by applying the same event and muon track selection criteria as described in Section~\ref{sec:selection}. Each muon track is extrapolated to the surface of the V0C detector, and the channel hit by the muon is determined. The average signals in any given channel $i$ are then obtained, separately for the events in which the muons hit that given channel ($\langle s_i^i\rangle$) or any other channel $j$ in the same V0C ring ($\langle s_j^i\rangle$). The signal corresponding to the presence of a muon track in the channel $i$ is then computed as $\langle  s_i^i\rangle-(\langle s_{j^\prime}^i\rangle+\langle s_{j^{\prime\prime}}^i\rangle)/2$, where indices $j^\prime$ and $j^{\prime\prime}$ correspond to channels located in the same V0C ring and at $\pm$90$\degree$ in azimuth with respect to the channel $i$. This procedure takes into account the fact that the multiplicity of the underlying event is affected by the presence of a muon at a given pseudorapidity. In addition, it avoids possible biases from short-range correlations from particle decays or recoil jets. Repeating the analysis with a higher muon trigger \pt threshold shows no significant contribution from muons produced within jets. Finally, the obtained muon track signals per channel are used in each dimuon event in order to subtract the contribution of the muons from the V0C signal in case one or both muons are within the V0C acceptance.

\subsection{Measurement of charged-particle multiplicity}
\label{sec:nch}

The measured V0C signal is transformed into primary charged-particle multiplicity with the help of PYTHIA~8.2~\cite{Sjostrand:2014zea} and EPOS3~\cite{Werner:2013tya} Monte-Carlo (MC) simulations. In order to take into account the residual detector aging effects discussed in Section~\ref{sec:mult_sel}, the V0C signal in the simulations is corrected in the same way as the data using the same global reference value of the mean V0C signal. The $z_{\rm vtx}$ correction is also applied. In Fig.~\ref{fig:V0CPythiaData}, the resulting simulated V0C signal distribution in the case of PYTHIA~8.2 MC is compared with the data.

The MC simulations are used to obtain the mean charged-particle multiplicity in bins of the corrected V0C signal, $\langle N_{\rm ch}\rangle^{\rm MC}_i$, where $i$ is the bin index. In the region of high multiplicity, where the binning leads to sizable statistical fluctuations, the mean multiplicity is fitted with a combination of exponential and linear functions of the V0C signal. The resulting mean multiplicities show a less-than-linear increase as a function of the V0C signal, especially at high V0C values. This is due to the fact that the V0C signal is proportional to the energy deposited by the charged particles in the thin scintillator layer and to the presence of the secondary particles produced in interactions of primary particles with the detector material in front of the V0C. These effects, coupled to the steeply falling charged-particle multiplicity distribution at high multiplicities (see Fig.~\ref{fig:V0CPythiaData}), lead to a situation in which the events with higher V0C signal get a higher relative contribution from collisions having lower charged-particle multiplicity, and upward fluctuations of the energy deposit in the scintillators or of the number of produced secondary particles. 

In the next step, the data V0C distribution in each multiplicity class and the integrated one are folded with $\langle N_{\rm ch}\rangle^{\rm MC}_i$ and the corresponding mean charged-particle multiplicities are calculated as $\sum_i{n^{\rm ev}_i \langle N_{\rm ch}\rangle^{\rm MC}_i}/\sum_i{n^{\rm ev}_i}$, where $n^{\rm ev}_i$ is the number of data events in bin $i$.

\subsection{\texorpdfstring{Measurement of \jpsi yields}{Measurement of J/psi yields}}
\label{sec:signal}
\begin{figure}[htb!]
    \centering
    \includegraphics[width=0.8\linewidth]{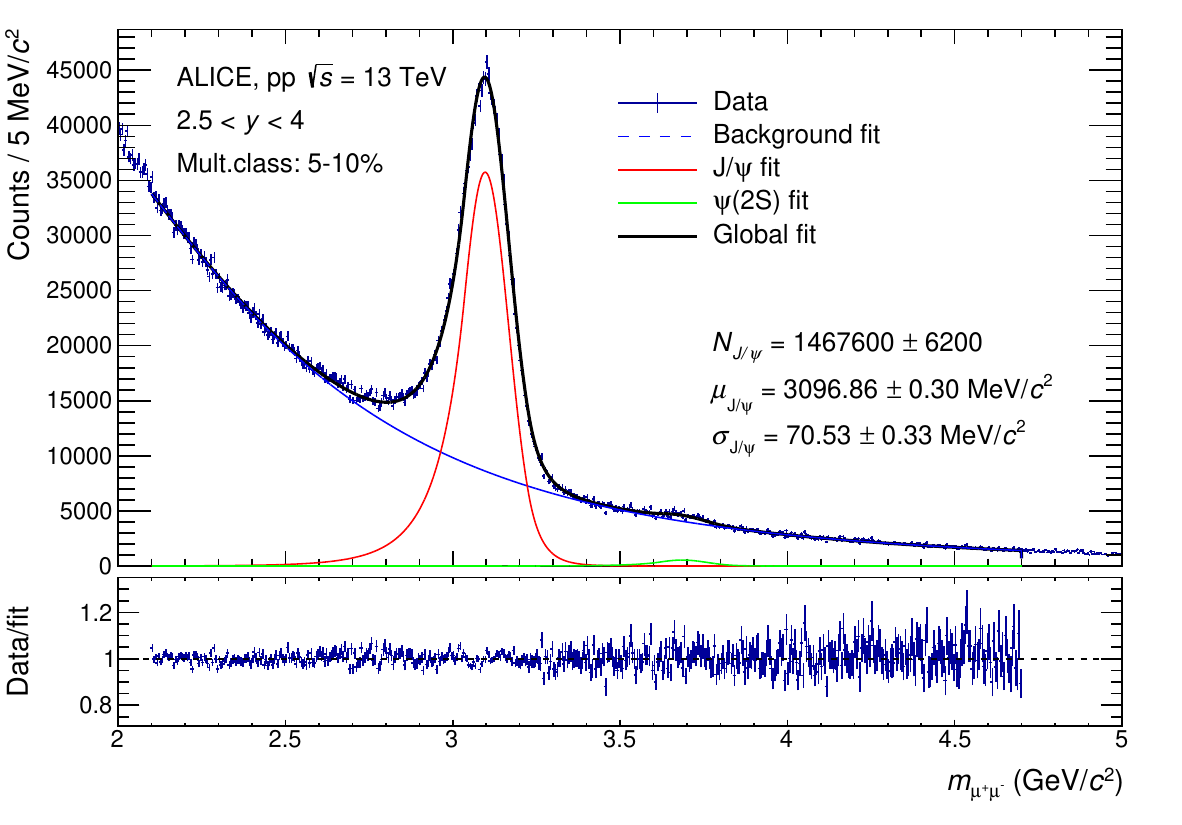}
    \caption{The dimuon invariant-mass distribution in the 5-10\% multiplicity class. Each dimuon is weighted with the inverse of the detector acceptance and efficiency as a function of the dimuon \pt and $y$. The data are fitted to a combination of double-sided Crystal Ball (CB2) functions for the \jpsi and $\psi(2S)$ signals, and a Variable-Width Gaussian function for the background. Only statistical uncertainties are shown. The bottom panel shows the ratio between the data and the fit.}
    \label{fig:jpsifit}
\end{figure}

The \jpsi yields are extracted via fits of the dimuon invariant-mass distributions in each multiplicity class~\cite{ResultsForward}. Each dimuon is weighted with the inverse of the detector acceptance and efficiency as a function of the dimuon \pt and $y$. The detector acceptance and efficiency are obtained by means of a dedicated $J/\psi\to\mu^+\mu^-$ MC simulation. The same values are used in all multiplicity classes since the detector acceptance and efficiency are not affected by the detector occupancy in pp collisions~\cite{ALICE:2022gpu}. The \jpsi signal is fitted with a double-sided Crystal-Ball (CB2) function~\cite{ALICE:2015ikg} with tail parameters extracted from either data or MC. The $\psi(2S)$ signal is also fitted with a CB2 function, but with parameters which are bind to those of the \jpsi CB2 function as described in Ref.~\cite{ResultsForward}.
The underlying dimuon background is fitted with either Variable-Width Gaussian function or a product of a second order polynomial and an exponential function~\cite{ALICE:2015ikg}. The fit is performed in the invariant-mass range of either 2.3 $<$ $m_{\mu\mu}$ $<$ 4.9 GeV/$c^2$ or 2.1 $<$ $m_{\mu\mu}$ $<$ 4.7 GeV/$c^2$. All the possible 8 fit configurations (2 sets of CB2 tail parameters, 2 background fit functions and 2 invariant-mass fit ranges) are considered. A typical fit example is shown in Fig.~\ref{fig:jpsifit}.

The \jpsi yields per collision are calculated as the ratios of the number of \jpsi obtained from the above fits and the number of corresponding collisions in the minimum-bias data set. This is done for each multiplicity class and for the integrated data set. The nominal values of the relative \jpsi yields and their statistical uncertainties are calculated as the averages of the  relative yields and their statistical uncertainties obtained from all the fit configurations discussed above.

\subsection{\texorpdfstring{Corrections of multiplicity and \jpsi yields}{Corrections of multiplicity and J/psi yields}}
\label{sec:corr}
As mentioned in Section~\ref{sec:intro}, the measurement is performed for 'INEL$>$0' event class, which requires corrections to the charged-particle multiplicity and the \jpsi yields. The corrections are obtained and applied following the approach in Refs.~\cite{ResultsForward,ALICE:2022gpu}. They are obtained separately in each multiplicity class and in the integrated data set, most of them with the help of a PYTHIA~8.2 MC simulation. 

The charged-particle multiplicity is corrected for the trigger and event selection efficiency. The value of the correction factor at the level of relative multiplicity is about 0.97 in the lowest multiplicity class and between 1.04 and 1.05 in the other multiplicity classes.

Three corrections are applied to the relative \jpsi yields. The corrections are related to the number of collisions used to calculate the \jpsi yields per collision. The first one is defined as a ratio of the number of triggered and selected (excluding vertex quality selection) 'INEL$>$0' collisions and the total number of 'INEL$>$0' collisions. The correction values range from 0.92 in the lowest multiplicity class up to 1.06 in the highest multiplicity class. The second correction is defined as a ratio of the number of triggered and selected collisions, including and excluding the vertex quality selection. It ranges from 0.94 in the lowest multiplicity class up to 1.04 in the highest multiplicity class. The third correction is introduced in order to take into account the contamination from 'INEL$=$0' collisions (inelastic collisions having no charged particles in the range $|\eta|$~$<$~1) in the sample of triggered and selected collisions. The correction is 1.03 in the lowest multiplicity class and reaches about 0.98 in high multiplicity classes. The first and third corrections are derived with PYTHIA~8.2 MC simulation, whereas the second correction is calculated from the data.

\section{Systematic uncertainties}

The conversion of the V0C signal into charged-particle multiplicity relies on the chosen MC model, which may not describe the data accurately enough. To address this, the charged-particle multiplicity is obtained using both PYTHIA~8.2 and EPOS3 MC simulations. In addition, the simulated V0C signal is scaled by $\pm$10\% before 
the conversion of V0C amplitude into charged-particle multiplicity, as described in Section~\ref{sec:nch}. The reason behind this scaling is that while both data and MC are normalized to the same global reference value of the mean V0C signal, the simulated charged-particle multiplicity in PYTHIA~8.2 is found to be up to about 10\% higher with respect to the data within the measured pseudorapidity range $|\eta|<$1.8~\cite{ALICE:2015qqj}. The statistical uncertainties of the parameters of the exponential and linear functions which are employed to fit the mean multiplicity as a function of the V0C signal at high multiplicity (see Section~\ref{sec:nch}) have essentially no impact on the results and are neglected. The systematic uncertainty of the charged-particle multiplicity is obtained as the standard deviation of the results obtained with the two MC simulations and with the above scaling of the V0C signal.

Several sources of systematic uncertainties of the \jpsi yields are considered. The uncertainty related to the \jpsi signal extraction is calculated as the standard deviation of the results obtained by varying from all the fit configurations discussed in Section~\ref{sec:signal}. The variations are applied simultaneously in the fits of each multiplicity class and of the integrated data set.

The uncertainties related to the subtraction of the muons contribution from the V0C signal are calculated by using extreme values of the muon track signals per channel found in various data-taking periods.

The differences between the results with and without the application of the corrections described in Section~\ref{sec:corr} are conservatively assigned as the corresponding systematic uncertainties of the charged-particle multiplicity and the \jpsi yields.

\begin{table}[ht]
\centering
\caption{Systematic uncertainties of the relative charged-particle multiplicities and the relative \jpsi yields. The uncertainties vary within the indicated ranges depending on the multiplicity classes.}
\begin{tabular}[ht]{c|c}
    \hline
    Source & Systematic uncertainties, \% \\
    \hline
    Conversion of V0C signal into $N_{\rm ch}$ & 1.2 - 3.4\\
    Correction of $N_{\rm ch}$ for INEL$>$0 & 2.7 - 4.9\\
    \hline
    Total syst. uncertainty of $N_{\rm ch}/\langle N_{\rm ch}\rangle$ & 4.3 - 5.5 \\
    \hline
    \jpsi signal extraction & 0.3 - 0.8\\
    Subtraction of $\mu$'s signal in V0C & 0.3 - 6.2\\
    Trigger efficiency & 1.3 - 5.7\\
    Vertex quality selection & 0.5 - 6.0\\
    Contamination from INEL=0 & 0.4 - 5.3\\
    \hline
    Total syst. uncertainty of $N^{\jpsi}/\langle N^{\jpsi}\rangle$ & 2.1 - 11.6 \\
    \hline
\end{tabular}
\label{table:syst}
\end{table}%

The systematic uncertainties of the relative charged-particle multiplicity and the relative \jpsi yields are summarized in Table~\ref{table:syst}. The total systematic uncertainties are calculated as quadratic sum of the uncertainties from the different sources. It is important to note that all the systematic uncertainties, except the one related to the \jpsi signal extraction, are fully correlated across the multiplicity classes and therefore one can also consider the total systematic uncertainty as practically fully correlated across the multiplicity classes.

\section{Results and discussion}
\begin{figure}[!ht]
    \centering
    \includegraphics[width=0.7\linewidth]{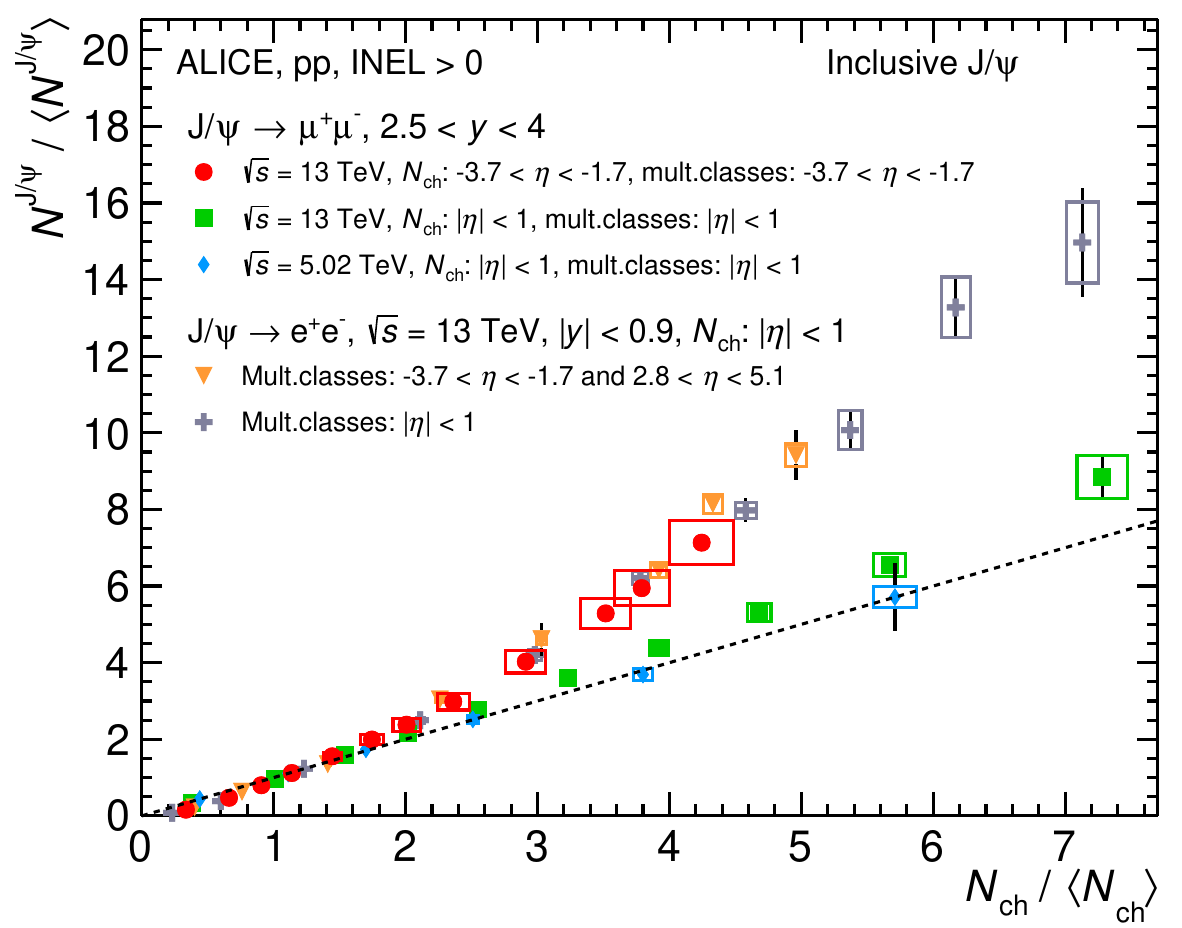}
    \caption{Forward and midrapidity \jpsi relative yields versus the relative charged-particle multiplicity. The relative forward \jpsi yields as a function of the relative forward charged-particle multiplicity are shown as red circles. The data of midrapidity \jpsi and forward rapidity \jpsi versus midrapidity multiplicity are taken from Refs.~\cite{ResultsMidrap} and~\cite{ResultsForward}, respectively. The error bars and boxes represent the statistical and systematic uncertainties, respectively. The dotted line represents the diagonal.}
    \label{fig:ResultMidRapAndMine}
\end{figure}

The relative forward \jpsi yields as a function of the relative forward charged-particle multiplicity are shown as red circles in Fig.~\ref{fig:ResultMidRapAndMine}. The observed dependence is steeper-than-linear. It is similar to the one of the midrapidity \jpsi yields as a function of the midrapidity charged-particle multiplicity (gray crosses and yellow triangles in Fig.~\ref{fig:ResultMidRapAndMine}) and is significantly steeper than the one of the forward \jpsi yields as a function of midrapidity charged-particle multiplicity (green squares and blue diamonds in Fig.~\ref{fig:ResultMidRapAndMine}). These results would advocate against any major difference in the production mechanisms of \jpsi at midrapidity versus forward rapidity. Instead, the close-to-linear dependence observed in the latter measurement could be due to weaker autocorrelation effects affecting the \jpsi yields and to the significantly wider spanned range in the relative multiplicity – resulting from the full overlap of the pseudorapidity ranges of the multiplicity estimator and the multiplicity measurement.

As it is described in Section~\ref{sec:selection}, the results presented here are obtained by excluding the contribution of the \jpsi muon daughters from the particle multiplicity estimation. In Ref.~\cite{PHENIX:2024dqs}, it is found that the inclusion of the daughters leads to a significant autocorrelation effect and has a major impact on the measured multiplicity dependence of the \jpsi yields. In the present analysis, the inclusion of the daughters has a much smaller effect on the \jpsi yields, reaching 3.4\% in the highest multiplicity class. In order to gain further understanding, a PYTHIA~8.3 Monash minimum-bias MC simulation is carried out at the generator level without transport through the detector. The produced \jpsi are forced to decay into two muons. It is found that the inclusion of the muon daughters leads to up to about 13\% higher \jpsi yields in the highest multiplicity class. The difference with respect to the data can be explained by two effects. First, the V0 detector delivers a signal proportional to the energy deposited by charged particles in its scintillator layer. While the muon daughter tracks produce signals close to that of a minimum-ionizing particle (MIP), most of the charged particles produced in the collision and reaching the V0 detector deposit significantly more energy than a MIP. Second, a sizable amount of secondary particles is produced in interactions of primary particles with the detector material in front of the V0C detector. This significantly increases the particle multiplicity within the acceptance of the detector and reduces the impact of the \jpsi daughter tracks on the multiplicity measurement.

The effect of the inclusion of the \jpsi decay daughters discussed above is orders of magnitude smaller compared to the measurement in Ref.~\cite{PHENIX:2024dqs}, this is presumably due to the much higher charged-particle multiplicities of the underlying event at the LHC energies compared to the RHIC ones. Finally, it is worth noting that the previous results at midrapidity at the same collision energy (gray crosses in Fig.~\ref{fig:ResultMidRapAndMine}) were obtained without excluding the contribution of the two electron daughters to the charged-particle multiplicity, which can affect the comparison with the present results.

\section{Comparison with models}

\begin{figure}[hbtp]
    \centering
    \includegraphics[width=0.7\textwidth]{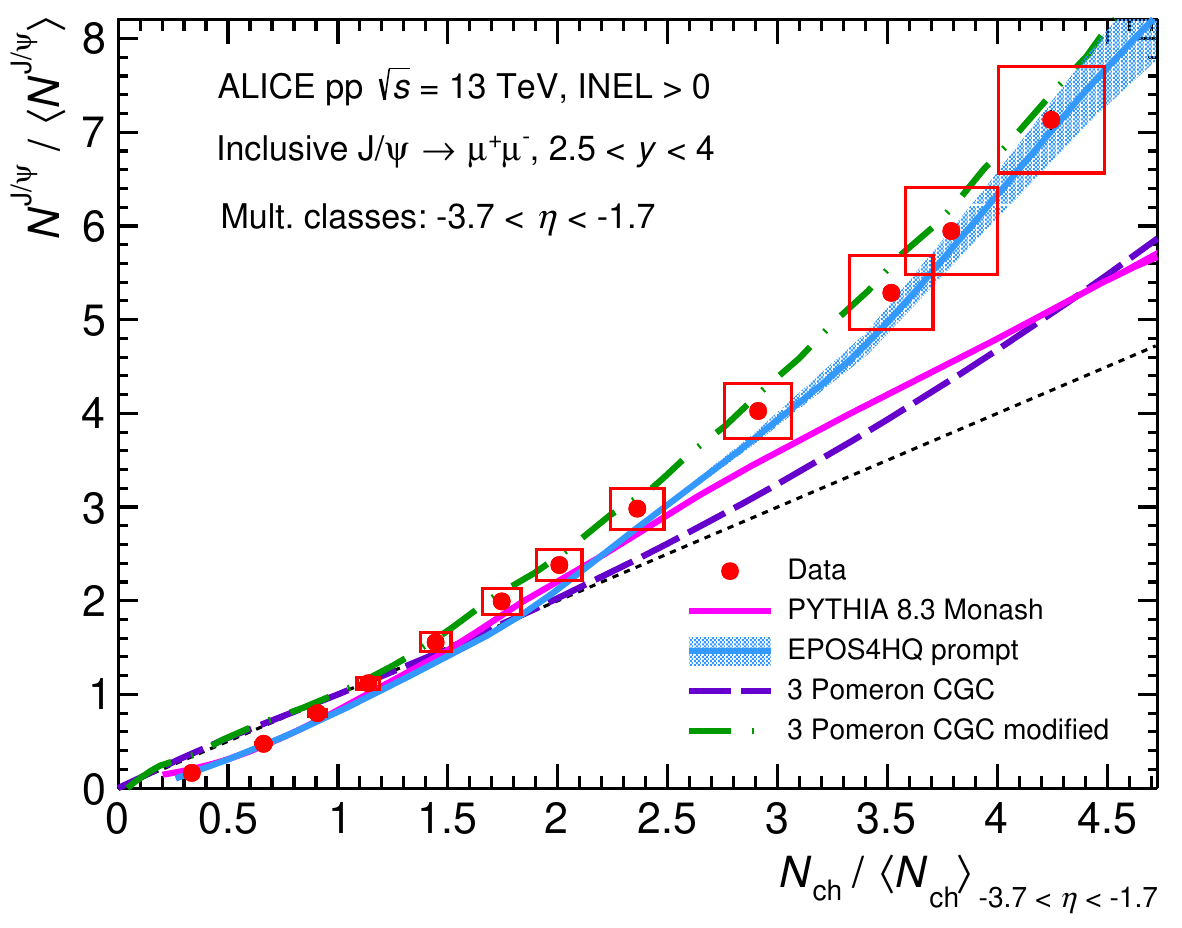}
    \caption{\jpsi relative yield as a function of the relative multiplicity in the V0C acceptance $-3.7<\eta<-1.7$ compared with PYTHIA~8.3~\cite{Bierlich:2022pfr}, EPOS4HQ~\cite{Zhao:2023ucp} and two 3-Pomeron CGC model predictions~\cite{Levin:2019fvb,Gotsman:2020ubn}. The dotted line represents the diagonal.}
    \label{fig:resultModelComp}
\end{figure}

In Fig.~\ref{fig:resultModelComp}, the results are compared with PYTHIA~8.3~\cite{Bierlich:2022pfr}, EPOS4HQ~\cite{Zhao:2023ucp}, and CGC model predictions~\cite{Levin:2019fvb,Gotsman:2020ubn}. PYTHIA~8.3 calculations consider both prompt and non-prompt \jpsi production, while EPOS4HQ and CGC models consider only prompt \jpsi production. The measured fraction of non-prompt \jpsi in pp collisions at $\sqrt{s}=13$ TeV is about 13\% at forward rapidity (2 $<$ $y$ $<$ 4.5) and integrated over \pt~\cite{LHCb:2015foc}. This sizeable relative contribution of non-prompt \jpsi and its possible multiplicity dependence have to be taken into account when interpreting the comparison between the data and EPOS4HQ and CGC model predictions.

EPOS4HQ is a version of the EPOS4 event generator. The EPOS4 generator features core-corona separation and formation of hydrodynamically evolving hot QCD medium in the core. The initial stage of the collision is described by means of multiple parallel partonic scatterings. The main difference with respect to previous versions of EPOS lies in the introduction of a dynamical saturation scale. The scale describes the saturation phenomena of very small momentum partons in the incoming protons and preserves the factorization of the initial-state parton distribution functions and parton-parton interaction cross section. In EPOS4HQ, heavy quarks, produced in initial partonic scatterings at the initial stage of the collision, can eventually interact with the medium and hadronize via coalescence~\cite{Werner:2023zvo}.
The charmonium formation is implemented via the Wigner density formalism~\cite{Song:2017phm}.

Both the considered 3-Pomeron CGC models describe the \jpsi production using the color-singlet contribution of NRQCD with 3-Pomeron fusion and CGC formalism for the incoming protons. The difference between the 3-Pomeron
CGC model developed in Ref.~\cite{Levin:2019fvb} and the modified 3-Pomeron CGC model described in Ref.~\cite{Gotsman:2020ubn} is that in the latter higher-order gluon interactions are included. Within the 3-Pomeron CGC models the steepness of the \jpsi yields as a function of the multiplicity is not affected by autocorrelation effects and is determined mainly by the level of overlap between the rapidity ranges of the measured \jpsi yields and charged-particle multiplicity. It is important to note that all the considered models do not include the \jpsi daughters in the counting of the charged-particle multiplicity in the events containing \jpsi or do not consider the \jpsi decays at all.

As discussed in Section~\ref{sec:intro}, in principle within the PYTHIA framework and modulo the contribution from non-prompt \jpsi production, the observed steeper-than-linear dependence could indicate that the CR mechanism has a major role in \jpsi production. However, as one can see in Fig.~\ref{fig:resultModelComp} PYTHIA~8.3 with enabled by default CR underestimates the results at intermediate and high multiplicities. It is worth noting that a similar level of underestimation is observed for the previous measurements of mid and forward rapidity \jpsi yields versus midrapidity multiplicity. This could also mean that the steeper-than-linear dependence is mainly driven by autocorrelation effects (e.g., from particles steaming from the b-quark fragmentation or associated-particle production mechanisms) and that these effects are underestimated in PYTHIA. Measurements of the \jpsi production in jets in pp collisions at the LHC~\cite{LHCb:2017llq,CMS:2021puf} seem to corroborate such a hypothesis as they show significantly higher associated-hadron production compared to PYTHIA. Nevertheless, no conclusions can be drawn from these measurements as they correspond to high-\pt~\jpsi production, whereas the \pt-integrated \jpsi yields discussed here are dominated by the low-\pt part of the \jpsi spectra.

The EPOS4HQ and the modified 3-Pomeron CGC model predictions are both consistent with the data, although these predictions consider only prompt \jpsi production while the data correspond to inclusive \jpsi yields. Interestingly, the former 3-Pomeron CGC model accurately describes the measurements of both midrapidity and forward rapidity \jpsi yields versus midrapidity multiplicity, but does not describe the results of the forward \jpsi yields versus forward multiplicity from
the present analysis. Previous versions of the EPOS event generator failed to describe the midrapidity and forward rapidity \jpsi yields as a function of midrapidity multiplicity, underestimating the yields at intermediate and high multiplicity~\cite{ResultsMidrap,ResultsForward}. This is no longer the case with the EPOS4HQ, which describes rather accurately the results presented here, up to the highest measured multiplicity class. The reasons likely lie in the introduction of the dynamical saturation scale and the implemented charmonium formation via coalescence.

\section{Conclusions}

The forward relative \jpsi yields are measured as a function of the relative charged-particle multiplicity at forward rapidity, in pp collisions at $\sqrt{s}$ = 13 TeV. The results show a steeper-than-linear dependence, similar to that of midrapidity \jpsi yields versus midrapidity multiplicity and significantly steeper than in the case of forward \jpsi yields versus midrapidity multiplicity at the same collision energy. This may indicate that the \jpsi production mechanisms at forward and midrapidity are similar and that the previously observed differences are likely due to effects related to autocorrelations in the measurement.

Although the observed steeper-than-linear dependence can be interpreted within the PYTHIA event generator as an indication of the important role of the Color Reconnection mechanism in the \jpsi production, the predictions calculated with PYTHIA~8.3 significantly underestimate the \jpsi yields at intermediate and high multiplicity. The results show a good agreement with the EPOS4HQ and the modified 3-Pomeron CGC models in the whole studied multiplicity range. It is worth noting that these models feature parton saturation in the incoming protons and consider only prompt \jpsi production.

These results need to be complemented by further studies before more firm conclusions can be drawn. For example, the measurement of midrapidity \jpsi yields versus forward multiplicity could be useful. 
Studying the \jpsi yields as a function of the multiplicity in the transverse-to-\jpsi region was proposed in Ref.~\cite{Weber:2018ddv} to further investigate the role of autocorrelation effects from associated hadroproduction. Recently, the event flattenicity was proposed as an additional tool in the selection of high-multiplicity collisions. The flattenicity is a measure of the isotropy of the distribution of the produced particles. It avoids the biases arising from local multiplicity fluctuations and autocorrelation effects and thus provides an unbiased selection of collisions with a high number of MPIs~\cite{Ortiz:2022mfv}. Studying the \jpsi yields as a function of both multiplicity and flattenicity can therefore be very useful to disentangle various effects involved in \jpsi production. Finally, the ALICE apparatus used in the ongoing LHC Run 3 features a new Muon Forward Tracker (MFT) detector~\cite{CERN-LHCC-2015-001} which will enable the separation of prompt and non-prompt \jpsi in the forward direction in future ALICE publications.


\newenvironment{acknowledgement}{\relax}{\relax}
\begin{acknowledgement}
\section*{Acknowledgements}
\input{fa_2025-03-14_Opt_C.tex}
\end{acknowledgement}

\bibliographystyle{utphys}   
\bibliography{bibliography}

\newpage
\appendix

%
%

\section{The ALICE Collaboration}
\label{app:collab}
\input{Alice_Authorlist_2025-03-14_Opt_C.tex}
\end{document}

%% file: commands.tex
%
\newcommand{\jpsi}{{\ensuremath{\rm J/\psi}}\xspace}

\newcommand{\pp}           {pp\xspace}
\newcommand{\ppbar}        {\mbox{$\mathrm {p\overline{p}}$}\xspace}
\newcommand{\XeXe}         {\mbox{Xe--Xe}\xspace}
\newcommand{\PbPb}         {\mbox{Pb--Pb}\xspace}
\newcommand{\pA}           {\mbox{pA}\xspace}
\newcommand{\pPb}          {\mbox{p--Pb}\xspace}
\newcommand{\AuAu}         {\mbox{Au--Au}\xspace}
\newcommand{\dAu}          {\mbox{d--Au}\xspace}

\newcommand{\s}            {\ensuremath{\sqrt{s}}\xspace}
\newcommand{\snn}          {\ensuremath{\sqrt{s_{\mathrm{NN}}}}\xspace}
\newcommand{\pt}           {\ensuremath{p_{\rm T}}\xspace}
\newcommand{\meanpt}       {$\langle p_{\mathrm{T}}\rangle$\xspace}
\newcommand{\ycms}         {\ensuremath{y_{\rm CMS}}\xspace}
\newcommand{\ylab}         {\ensuremath{y_{\rm lab}}\xspace}
\newcommand{\etarange}[1]  {\mbox{$\left | \eta \right |~<~#1$}}
\newcommand{\yrange}[1]    {\mbox{$\left | y \right |~<~#1$}}
\newcommand{\dndy}         {\ensuremath{\mathrm{d}N_\mathrm{ch}/\mathrm{d}y}\xspace}
\newcommand{\dndeta}       {\ensuremath{\mathrm{d}N_\mathrm{ch}/\mathrm{d}\eta}\xspace}
\newcommand{\avdndeta}     {\ensuremath{\langle\dndeta\rangle}\xspace}
\newcommand{\dNdy}         {\ensuremath{\mathrm{d}N_\mathrm{ch}/\mathrm{d}y}\xspace}
\newcommand{\Npart}        {\ensuremath{N_\mathrm{part}}\xspace}
\newcommand{\Ncoll}        {\ensuremath{N_\mathrm{coll}}\xspace}
\newcommand{\dEdx}         {\ensuremath{\textrm{d}E/\textrm{d}x}\xspace}
\newcommand{\RpPb}         {\ensuremath{R_{\rm pPb}}\xspace}

\newcommand{\nineH}        {$\sqrt{s}~=~0.9$~Te\kern-.1emV\xspace}
\newcommand{\seven}        {$\sqrt{s}~=~7$~Te\kern-.1emV\xspace}
\newcommand{\twoH}         {$\sqrt{s}~=~0.2$~Te\kern-.1emV\xspace}
\newcommand{\twosevensix}  {$\sqrt{s}~=~2.76$~Te\kern-.1emV\xspace}
\newcommand{\five}         {$\sqrt{s}~=~5.02$~Te\kern-.1emV\xspace}
\newcommand{\twosevensixnn}{$\sqrt{s_{\mathrm{NN}}}~=~2.76$~Te\kern-.1emV\xspace}
\newcommand{\fivenn}       {$\sqrt{s_{\mathrm{NN}}}~=~5.02$~Te\kern-.1emV\xspace}
\newcommand{\LT}           {L{\'e}vy-Tsallis\xspace}
\newcommand{\GeVc}         {Ge\kern-.1emV/$c$\xspace}
\newcommand{\MeVc}         {Me\kern-.1emV/$c$\xspace}
\newcommand{\TeV}          {Te\kern-.1emV\xspace}
\newcommand{\GeV}          {Ge\kern-.1emV\xspace}
\newcommand{\MeV}          {Me\kern-.1emV\xspace}
\newcommand{\GeVmass}      {Ge\kern-.2emV/$c^2$\xspace}
\newcommand{\MeVmass}      {Me\kern-.2emV/$c^2$\xspace}
\newcommand{\lumi}         {\ensuremath{\mathcal{L}}\xspace}

\newcommand{\ITS}          {\rm{ITS}\xspace}
\newcommand{\TOF}          {\rm{TOF}\xspace}
\newcommand{\ZDC}          {\rm{ZDC}\xspace}
\newcommand{\ZDCs}         {\rm{ZDCs}\xspace}
\newcommand{\ZNA}          {\rm{ZNA}\xspace}
\newcommand{\ZNC}          {\rm{ZNC}\xspace}
\newcommand{\SPD}          {\rm{SPD}\xspace}
\newcommand{\SDD}          {\rm{SDD}\xspace}
\newcommand{\SSD}          {\rm{SSD}\xspace}
\newcommand{\TPC}          {\rm{TPC}\xspace}
\newcommand{\TRD}          {\rm{TRD}\xspace}
\newcommand{\VZERO}        {\rm{V0}\xspace}
\newcommand{\VZEROA}       {\rm{V0A}\xspace}
\newcommand{\VZEROC}       {\rm{V0C}\xspace}
\newcommand{\Vdecay} 	   {\ensuremath{V^{0}}\xspace}

\newcommand{\ee}           {\ensuremath{e^{+}e^{-}}} 
\newcommand{\pip}          {\ensuremath{\pi^{+}}\xspace}
\newcommand{\pim}          {\ensuremath{\pi^{-}}\xspace}
\newcommand{\kap}          {\ensuremath{\rm{K}^{+}}\xspace}
\newcommand{\kam}          {\ensuremath{\rm{K}^{-}}\xspace}
\newcommand{\pbar}         {\ensuremath{\rm\overline{p}}\xspace}
\newcommand{\kzero}        {\ensuremath{{\rm K}^{0}_{\rm{S}}}\xspace}
\newcommand{\lmb}          {\ensuremath{\Lambda}\xspace}
\newcommand{\almb}         {\ensuremath{\overline{\Lambda}}\xspace}
\newcommand{\Om}           {\ensuremath{\Omega^-}\xspace}
\newcommand{\Mo}           {\ensuremath{\overline{\Omega}^+}\xspace}
\newcommand{\X}            {\ensuremath{\Xi^-}\xspace}
\newcommand{\Ix}           {\ensuremath{\overline{\Xi}^+}\xspace}
\newcommand{\Xis}          {\ensuremath{\Xi^{\pm}}\xspace}
\newcommand{\Oms}          {\ensuremath{\Omega^{\pm}}\xspace}
\newcommand{\degree}       {\ensuremath{^{\rm o}}\xspace}

%% file: fa_2025-03-14_Opt_C.tex

The ALICE Collaboration would like to thank all its engineers and technicians for their invaluable contributions to the construction of the experiment and the CERN accelerator teams for the outstanding performance of the LHC complex.
The ALICE Collaboration gratefully acknowledges the resources and support provided by all Grid centres and the Worldwide LHC Computing Grid (WLCG) collaboration.
The ALICE Collaboration acknowledges the following funding agencies for their support in building and running the ALICE detector:
A. I. Alikhanyan National Science Laboratory (Yerevan Physics Institute) Foundation (ANSL), State Committee of Science and World Federation of Scientists (WFS), Armenia;
Austrian Academy of Sciences, Austrian Science Fund (FWF): [M 2467-N36] and Nationalstiftung f\"{u}r Forschung, Technologie und Entwicklung, Austria;
Ministry of Communications and High Technologies, National Nuclear Research Center, Azerbaijan;
Conselho Nacional de Desenvolvimento Cient\'{\i}fico e Tecnol\'{o}gico (CNPq), Financiadora de Estudos e Projetos (Finep), Funda\c{c}\~{a}o de Amparo \`{a} Pesquisa do Estado de S\~{a}o Paulo (FAPESP) and Universidade Federal do Rio Grande do Sul (UFRGS), Brazil;
Bulgarian Ministry of Education and Science, within the National Roadmap for Research Infrastructures 2020-2027 (object CERN), Bulgaria;
Ministry of Education of China (MOEC) , Ministry of Science \& Technology of China (MSTC) and National Natural Science Foundation of China (NSFC), China;
Ministry of Science and Education and Croatian Science Foundation, Croatia;
Centro de Aplicaciones Tecnol\'{o}gicas y Desarrollo Nuclear (CEADEN), Cubaenerg\'{\i}a, Cuba;
Ministry of Education, Youth and Sports of the Czech Republic, Czech Republic;
The Danish Council for Independent Research | Natural Sciences, the VILLUM FONDEN and Danish National Research Foundation (DNRF), Denmark;
Helsinki Institute of Physics (HIP), Finland;
Commissariat \`{a} l'Energie Atomique (CEA) and Institut National de Physique Nucl\'{e}aire et de Physique des Particules (IN2P3) and Centre National de la Recherche Scientifique (CNRS), France;
Bundesministerium f\"{u}r Bildung und Forschung (BMBF) and GSI Helmholtzzentrum f\"{u}r Schwerionenforschung GmbH, Germany;
General Secretariat for Research and Technology, Ministry of Education, Research and Religions, Greece;
National Research, Development and Innovation Office, Hungary;
Department of Atomic Energy Government of India (DAE), Department of Science and Technology, Government of India (DST), University Grants Commission, Government of India (UGC) and Council of Scientific and Industrial Research (CSIR), India;
National Research and Innovation Agency - BRIN, Indonesia;
Istituto Nazionale di Fisica Nucleare (INFN), Italy;
Japanese Ministry of Education, Culture, Sports, Science and Technology (MEXT) and Japan Society for the Promotion of Science (JSPS) KAKENHI, Japan;
Consejo Nacional de Ciencia (CONACYT) y Tecnolog\'{i}a, through Fondo de Cooperaci\'{o}n Internacional en Ciencia y Tecnolog\'{i}a (FONCICYT) and Direcci\'{o}n General de Asuntos del Personal Academico (DGAPA), Mexico;
Nederlandse Organisatie voor Wetenschappelijk Onderzoek (NWO), Netherlands;
The Research Council of Norway, Norway;
Pontificia Universidad Cat\'{o}lica del Per\'{u}, Peru;
Ministry of Science and Higher Education, National Science Centre and WUT ID-UB, Poland;
Korea Institute of Science and Technology Information and National Research Foundation of Korea (NRF), Republic of Korea;
Ministry of Education and Scientific Research, Institute of Atomic Physics, Ministry of Research and Innovation and Institute of Atomic Physics and Universitatea Nationala de Stiinta si Tehnologie Politehnica Bucuresti, Romania;
Ministerstvo skolstva, vyskumu, vyvoja a mladeze SR, Slovakia;
National Research Foundation of South Africa, South Africa;
Swedish Research Council (VR) and Knut \& Alice Wallenberg Foundation (KAW), Sweden;
European Organization for Nuclear Research, Switzerland;
Suranaree University of Technology (SUT), National Science and Technology Development Agency (NSTDA) and National Science, Research and Innovation Fund (NSRF via PMU-B B05F650021), Thailand;
Turkish Energy, Nuclear and Mineral Research Agency (TENMAK), Turkey;
National Academy of  Sciences of Ukraine, Ukraine;
Science and Technology Facilities Council (STFC), United Kingdom;
National Science Foundation of the United States of America (NSF) and United States Department of Energy, Office of Nuclear Physics (DOE NP), United States of America.
In addition, individual groups or members have received support from:
Czech Science Foundation (grant no. 23-07499S), Czech Republic;
FORTE project, reg.\ no.\ CZ.02.01.01/00/22\_008/0004632, Czech Republic, co-funded by the European Union, Czech Republic;
European Research Council (grant no. 950692), European Union;
Deutsche Forschungs Gemeinschaft (DFG, German Research Foundation) ``Neutrinos and Dark Matter in Astro- and Particle Physics'' (grant no. SFB 1258), Germany;
ICSC - National Research Center for High Performance Computing, Big Data and Quantum Computing and FAIR - Future Artificial Intelligence Research, funded by the NextGenerationEU program (Italy).

%% file: Alice_Authorlist_2025-03-14_Opt_C.tex
\begin{flushleft} 
\small

S.~Acharya\,\orcidlink{0000-0002-9213-5329}\,$^{\rm 50}$, 
A.~Agarwal$^{\rm 133}$, 
G.~Aglieri Rinella\,\orcidlink{0000-0002-9611-3696}\,$^{\rm 32}$, 
L.~Aglietta\,\orcidlink{0009-0003-0763-6802}\,$^{\rm 24}$, 
M.~Agnello\,\orcidlink{0000-0002-0760-5075}\,$^{\rm 29}$, 
N.~Agrawal\,\orcidlink{0000-0003-0348-9836}\,$^{\rm 25}$, 
Z.~Ahammed\,\orcidlink{0000-0001-5241-7412}\,$^{\rm 133}$, 
S.~Ahmad\,\orcidlink{0000-0003-0497-5705}\,$^{\rm 15}$, 
S.U.~Ahn\,\orcidlink{0000-0001-8847-489X}\,$^{\rm 71}$, 
I.~Ahuja\,\orcidlink{0000-0002-4417-1392}\,$^{\rm 36}$
% ...

\section*{Affiliation Notes}

$^{\rm I}$ Deceased\\
$^{\rm II}$ Also at: Max-Planck-Institut fur Physik, Munich, Germany\\
$^{\rm III}$ Also at: Italian National Agency for New Technologies, Energy and Sustainable Economic Development (ENEA), Bologna, Italy\\
$^{\rm IV}$ Also at: Instituto de Fisica da Universidade de Sao Paulo\\
$^{\rm V}$ Also at: Dipartimento DET del Politecnico di Torino, Turin, Italy\\
$^{\rm VI}$ Also at: Department of Applied Physics, Aligarh Muslim University, Aligarh, India\\
$^{\rm VII}$ Also at: Institute of Theoretical Physics, University of Wroclaw, Poland\\
$^{\rm VIII}$ Also at: Facultad de Ciencias, Universidad Nacional Aut\'{o}noma de M\'{e}xico, Mexico City, Mexico\\

\section*{Collaboration Institutes}

$^{1}$ A.I. Alikhanyan National Science Laboratory (Yerevan Physics Institute) Foundation, Yerevan, Armenia\\
$^{2}$ AGH University of Krakow, Cracow, Poland\\
$^{3}$ Bogolyubov Institute for Theoretical Physics, National Academy of Sciences of Ukraine, Kiev, Ukraine\\
$^{4}$ Bose Institute, Department of Physics  and Centre for Astroparticle Physics and Space Science (CAPSS), Kolkata, India\\
$^{5}$ California Polytechnic State University, San Luis Obispo, California, United States\\
$^{6}$ Central China Normal University, Wuhan, China\\
$^{7}$ Centro de Aplicaciones Tecnol\'{o}gicas y Desarrollo Nuclear (CEADEN), Havana, Cuba\\
$^{8}$ Centro de Investigaci\'{o}n y de Estudios Avanzados (CINVESTAV), Mexico City and M\'{e}rida, Mexico\\
$^{9}$ Chicago State University, Chicago, Illinois, United States\\
$^{10}$ China Nuclear Data Center, China Institute of Atomic Energy, Beijing, China\\
$^{11}$ China University of Geosciences, Wuhan, China\\
$^{12}$ Chungbuk National University, Cheongju, Republic of Korea\\
$^{13}$ Comenius University Bratislava, Faculty of Mathematics, Physics and Informatics, Bratislava, Slovak Republic\\
$^{14}$ Creighton University, Omaha, Nebraska, United States\\
$^{15}$ Department of Physics, Aligarh Muslim University, Aligarh, India\\
$^{16}$ Department of Physics, Pusan National University, Pusan, Republic of Korea\\
$^{17}$ Department of Physics, Sejong University, Seoul, Republic of Korea\\
$^{18}$ Department of Physics, University of California, Berkeley, California, United States\\
$^{19}$ Department of Physics, University of Oslo, Oslo, Norway\\
$^{20}$ Department of Physics and Technology, University of Bergen, Bergen, Norway\\
$^{21}$ Dipartimento di Fisica, Universit\`{a} di Pavia, Pavia, Italy\\
$^{22}$ Dipartimento di Fisica dell'Universit\`{a} and Sezione INFN, Cagliari, Italy\\
$^{23}$ Dipartimento di Fisica dell'Universit\`{a} and Sezione INFN, Trieste, Italy\\
$^{24}$ Dipartimento di Fisica dell'Universit\`{a} and Sezione INFN, Turin, Italy\\
$^{25}$ Dipartimento di Fisica e Astronomia dell'Universit\`{a} and Sezione INFN, Bologna, Italy\\
$^{26}$ Dipartimento di Fisica e Astronomia dell'Universit\`{a} and Sezione INFN, Catania, Italy\\
$^{27}$ Dipartimento di Fisica e Astronomia dell'Universit\`{a} and Sezione INFN, Padova, Italy\\
$^{28}$ Dipartimento di Fisica `E.R.~Caianiello' dell'Universit\`{a} and Gruppo Collegato INFN, Salerno, Italy\\
$^{29}$ Dipartimento DISAT del Politecnico and Sezione INFN, Turin, Italy\\
$^{30}$ Dipartimento di Scienze MIFT, Universit\`{a} di Messina, Messina, Italy\\
$^{31}$ Dipartimento Interateneo di Fisica `M.~Merlin' and Sezione INFN, Bari, Italy\\
$^{32}$ European Organization for Nuclear Research (CERN), Geneva, Switzerland\\
$^{33}$ Faculty of Electrical Engineering, Mechanical Engineering and Naval Architecture, University of Split, Split, Croatia\\
$^{34}$ Faculty of Nuclear Sciences and Physical Engineering, Czech Technical University in Prague, Prague, Czech Republic\\
$^{35}$ Faculty of Physics, Sofia University, Sofia, Bulgaria\\
$^{36}$ Faculty of Science, P.J.~\v{S}af\'{a}rik University, Ko\v{s}ice, Slovak Republic\\
$^{37}$ Faculty of Technology, Environmental and Social Sciences, Bergen, Norway\\
$^{38}$ Frankfurt Institute for Advanced Studies, Johann Wolfgang Goethe-Universit\"{a}t Frankfurt, Frankfurt, Germany\\
$^{39}$ Fudan University, Shanghai, China\\
$^{40}$ Gangneung-Wonju National University, Gangneung, Republic of Korea\\
$^{41}$ Gauhati University, Department of Physics, Guwahati, India\\
$^{42}$ Helmholtz-Institut f\"{u}r Strahlen- und Kernphysik, Rheinische Friedrich-Wilhelms-Universit\"{a}t Bonn, Bonn, Germany\\
$^{43}$ Helsinki Institute of Physics (HIP), Helsinki, Finland\\
$^{44}$ High Energy Physics Group,  Universidad Aut\'{o}noma de Puebla, Puebla, Mexico\\
$^{45}$ Horia Hulubei National Institute of Physics and Nuclear Engineering, Bucharest, Romania\\
$^{46}$ HUN-REN Wigner Research Centre for Physics, Budapest, Hungary\\
$^{47}$ Indian Institute of Technology Bombay (IIT), Mumbai, India\\
$^{48}$ Indian Institute of Technology Indore, Indore, India\\
$^{49}$ INFN, Laboratori Nazionali di Frascati, Frascati, Italy\\
$^{50}$ INFN, Sezione di Bari, Bari, Italy\\
$^{51}$ INFN, Sezione di Bologna, Bologna, Italy\\
$^{52}$ INFN, Sezione di Cagliari, Cagliari, Italy\\
$^{53}$ INFN, Sezione di Catania, Catania, Italy\\
$^{54}$ INFN, Sezione di Padova, Padova, Italy\\
$^{55}$ INFN, Sezione di Pavia, Pavia, Italy\\
$^{56}$ INFN, Sezione di Torino, Turin, Italy\\
$^{57}$ INFN, Sezione di Trieste, Trieste, Italy\\
$^{58}$ Inha University, Incheon, Republic of Korea\\
$^{59}$ Institute for Gravitational and Subatomic Physics (GRASP), Utrecht University/Nikhef, Utrecht, Netherlands\\
$^{60}$ Institute of Experimental Physics, Slovak Academy of Sciences, Ko\v{s}ice, Slovak Republic\\
$^{61}$ Institute of Physics, Homi Bhabha National Institute, Bhubaneswar, India\\
$^{62}$ Institute of Physics of the Czech Academy of Sciences, Prague, Czech Republic\\
$^{63}$ Institute of Space Science (ISS), Bucharest, Romania\\
$^{64}$ Institut f\"{u}r Kernphysik, Johann Wolfgang Goethe-Universit\"{a}t Frankfurt, Frankfurt, Germany\\
$^{65}$ Instituto de Ciencias Nucleares, Universidad Nacional Aut\'{o}noma de M\'{e}xico, Mexico City, Mexico\\
$^{66}$ Instituto de F\'{i}sica, Universidade Federal do Rio Grande do Sul (UFRGS), Porto Alegre, Brazil\\
$^{67}$ Instituto de F\'{\i}sica, Universidad Nacional Aut\'{o}noma de M\'{e}xico, Mexico City, Mexico\\
$^{68}$ iThemba LABS, National Research Foundation, Somerset West, South Africa\\
$^{69}$ Jeonbuk National University, Jeonju, Republic of Korea\\
$^{70}$ Johann-Wolfgang-Goethe Universit\"{a}t Frankfurt Institut f\"{u}r Informatik, Fachbereich Informatik und Mathematik, Frankfurt, Germany\\
$^{71}$ Korea Institute of Science and Technology Information, Daejeon, Republic of Korea\\
$^{72}$ Laboratoire de Physique Subatomique et de Cosmologie, Universit\'{e} Grenoble-Alpes, CNRS-IN2P3, Grenoble, France\\
$^{73}$ Lawrence Berkeley National Laboratory, Berkeley, California, United States\\
$^{74}$ Lund University Department of Physics, Division of Particle Physics, Lund, Sweden\\
$^{75}$ Nagasaki Institute of Applied Science, Nagasaki, Japan\\
$^{76}$ Nara Women{'}s University (NWU), Nara, Japan\\
$^{77}$ National and Kapodistrian University of Athens, School of Science, Department of Physics , Athens, Greece\\
$^{78}$ National Centre for Nuclear Research, Warsaw, Poland\\
$^{79}$ National Institute of Science Education and Research, Homi Bhabha National Institute, Jatni, India\\
$^{80}$ National Nuclear Research Center, Baku, Azerbaijan\\
$^{81}$ National Research and Innovation Agency - BRIN, Jakarta, Indonesia\\
$^{82}$ Niels Bohr Institute, University of Copenhagen, Copenhagen, Denmark\\
$^{83}$ Nikhef, National institute for subatomic physics, Amsterdam, Netherlands\\
$^{84}$ Nuclear Physics Group, STFC Daresbury Laboratory, Daresbury, United Kingdom\\
$^{85}$ Nuclear Physics Institute of the Czech Academy of Sciences, Husinec-\v{R}e\v{z}, Czech Republic\\
$^{86}$ Oak Ridge National Laboratory, Oak Ridge, Tennessee, United States\\
$^{87}$ Ohio State University, Columbus, Ohio, United States\\
$^{88}$ Physics department, Faculty of science, University of Zagreb, Zagreb, Croatia\\
$^{89}$ Physics Department, Panjab University, Chandigarh, India\\
$^{90}$ Physics Department, University of Jammu, Jammu, India\\
$^{91}$ Physics Program and International Institute for Sustainability with Knotted Chiral Meta Matter (WPI-SKCM$^{2}$), Hiroshima University, Hiroshima, Japan\\
$^{92}$ Physikalisches Institut, Eberhard-Karls-Universit\"{a}t T\"{u}bingen, T\"{u}bingen, Germany\\
$^{93}$ Physikalisches Institut, Ruprecht-Karls-Universit\"{a}t Heidelberg, Heidelberg, Germany\\
$^{94}$ Physik Department, Technische Universit\"{a}t M\"{u}nchen, Munich, Germany\\
$^{95}$ Politecnico di Bari and Sezione INFN, Bari, Italy\\
$^{96}$ Research Division and ExtreMe Matter Institute EMMI, GSI Helmholtzzentrum f\"ur Schwerionenforschung GmbH, Darmstadt, Germany\\
$^{97}$ Saga University, Saga, Japan\\
$^{98}$ Saha Institute of Nuclear Physics, Homi Bhabha National Institute, Kolkata, India\\
$^{99}$ School of Physics and Astronomy, University of Birmingham, Birmingham, United Kingdom\\
$^{100}$ Secci\'{o}n F\'{\i}sica, Departamento de Ciencias, Pontificia Universidad Cat\'{o}lica del Per\'{u}, Lima, Peru\\
$^{101}$ Stefan Meyer Institut f\"{u}r Subatomare Physik (SMI), Vienna, Austria\\
$^{102}$ SUBATECH, IMT Atlantique, Nantes Universit\'{e}, CNRS-IN2P3, Nantes, France\\
$^{103}$ Sungkyunkwan University, Suwon City, Republic of Korea\\
$^{104}$ Suranaree University of Technology, Nakhon Ratchasima, Thailand\\
$^{105}$ Technical University of Ko\v{s}ice, Ko\v{s}ice, Slovak Republic\\
$^{106}$ The Henryk Niewodniczanski Institute of Nuclear Physics, Polish Academy of Sciences, Cracow, Poland\\
$^{107}$ The University of Texas at Austin, Austin, Texas, United States\\
$^{108}$ Universidad Aut\'{o}noma de Sinaloa, Culiac\'{a}n, Mexico\\
$^{109}$ Universidade de S\~{a}o Paulo (USP), S\~{a}o Paulo, Brazil\\
$^{110}$ Universidade Estadual de Campinas (UNICAMP), Campinas, Brazil\\
$^{111}$ Universidade Federal do ABC, Santo Andre, Brazil\\
$^{112}$ Universitatea Nationala de Stiinta si Tehnologie Politehnica Bucuresti, Bucharest, Romania\\
$^{113}$ University of Derby, Derby, United Kingdom\\
$^{114}$ University of Houston, Houston, Texas, United States\\
$^{115}$ University of Jyv\"{a}skyl\"{a}, Jyv\"{a}skyl\"{a}, Finland\\
$^{116}$ University of Kansas, Lawrence, Kansas, United States\\
$^{117}$ University of Liverpool, Liverpool, United Kingdom\\
$^{118}$ University of Science and Technology of China, Hefei, China\\
$^{119}$ University of South-Eastern Norway, Kongsberg, Norway\\
$^{120}$ University of Tennessee, Knoxville, Tennessee, United States\\
$^{121}$ University of the Witwatersrand, Johannesburg, South Africa\\
$^{122}$ University of Tokyo, Tokyo, Japan\\
$^{123}$ University of Tsukuba, Tsukuba, Japan\\
$^{124}$ Universit\"{a}t M\"{u}nster, Institut f\"{u}r Kernphysik, M\"{u}nster, Germany\\
$^{125}$ Universit\'{e} Clermont Auvergne, CNRS/IN2P3, LPC, Clermont-Ferrand, France\\
$^{126}$ Universit\'{e} de Lyon, CNRS/IN2P3, Institut de Physique des 2 Infinis de Lyon, Lyon, France\\
$^{127}$ Universit\'{e} de Strasbourg, CNRS, IPHC UMR 7178, F-67000 Strasbourg, France, Strasbourg, France\\
$^{128}$ Universit\'{e} Paris-Saclay, Centre d'Etudes de Saclay (CEA), IRFU, D\'{e}partment de Physique Nucl\'{e}aire (DPhN), Saclay, France\\
$^{129}$ Universit\'{e}  Paris-Saclay, CNRS/IN2P3, IJCLab, Orsay, France\\
$^{130}$ Universit\`{a} degli Studi di Foggia, Foggia, Italy\\
$^{131}$ Universit\`{a} del Piemonte Orientale, Vercelli, Italy\\
$^{132}$ Universit\`{a} di Brescia, Brescia, Italy\\
$^{133}$ Variable Energy Cyclotron Centre, Homi Bhabha National Institute, Kolkata, India\\
$^{134}$ Warsaw University of Technology, Warsaw, Poland\\
$^{135}$ Wayne State University, Detroit, Michigan, United States\\
$^{136}$ Yale University, New Haven, Connecticut, United States\\
$^{137}$ Yildiz Technical University, Istanbul, Turkey\\
$^{138}$ Yonsei University, Seoul, Republic of Korea\\
$^{139}$ Affiliated with an institute formerly covered by a cooperation agreement with CERN\\
$^{140}$ Affiliated with an international laboratory covered by a cooperation agreement with CERN.\\

\end{flushleft} 